\documentclass[reprint,aps,prl,superscriptaddress,longbibliography]{revtex4-1}
\usepackage[latin9]{inputenc}
\usepackage{color}
\usepackage{amsmath}
\usepackage{amssymb}
\usepackage{stmaryrd}
\usepackage{graphicx}
\usepackage{romannum}

\usepackage[unicode=true,
 bookmarks=true,bookmarksnumbered=false,bookmarksopen=false,
 breaklinks=false,pdfborder={0 0 1},backref=false,colorlinks=true]
 {hyperref}
\hypersetup{
 linkcolor=magenta, urlcolor=blue, citecolor=blue, pdfstartview={FitH}, hyperfootnotes=true, unicode=true}
 
\makeatletter
\@ifundefined{textcolor}{}
{%
 \definecolor{BLACK}{gray}{0}
 \definecolor{WHITE}{gray}{1}
 \definecolor{RED}{rgb}{1,0,0}
 \definecolor{GREEN}{rgb}{0,1,0}
 \definecolor{BLUE}{rgb}{0,0,1}
 \definecolor{CYAN}{cmyk}{1,0,0,0}
 \definecolor{MAGENTA}{cmyk}{0,1,0,0}
 \definecolor{YELLOW}{cmyk}{0,0,1,0}
}

\newcommand{\doublewidetilde}[1]{{%
  \mathpalette\double@widetilde{#1}%
}}
\newcommand{\double@widetilde}[2]{%
  \sbox\z@{$\m@th#1\widetilde{#2}$}%
  \ht\z@=.9\ht\z@
  \widetilde{\box\z@}%
}

\definecolor{darkgreen}{rgb}{0.0, 0.6, 0.13}

\usepackage{amsfonts}\usepackage{tabularx}\usepackage{dcolumn}\usepackage{bm}\usepackage{graphicx}\usepackage{epstopdf}

\hypersetup{urlcolor=blue}

\usepackage{tensor}
\usepackage{braket}

\usepackage[capitalise,compress]{cleveref}

\makeatother
\begin{document}
\title{Realizing and Probing  Baryonic Excitations in Rydberg Atom  Arrays}

\author{Fangli Liu}\thanks{These authors contributed equally to this work. Correspondence to: fliu1235@umd.edu (F.L.); spwhitsitt@gmail.com (S.W.)}
  \affiliation{Joint Quantum Institute and Joint Center for Quantum Information and Computer Science, NIST/University of Maryland, College Park, MD, 20742, USA}

 \author{Seth Whitsitt}
 \thanks{These authors contributed equally to this work. Correspondence to: fliu1235@umd.edu (F.L.); spwhitsitt@gmail.com (S.W.)}

  \affiliation{Joint Quantum Institute and Joint Center for Quantum Information and Computer Science, NIST/University of Maryland, College Park, MD, 20742, USA}
 
 \author{ Przemyslaw Bienias}
  \affiliation{Joint Quantum Institute and Joint Center for Quantum Information and Computer Science, NIST/University of Maryland, College Park, MD, 20742, USA}

 \author{ Rex Lundgren}
  \affiliation{Joint Quantum Institute and Joint Center for Quantum Information and Computer Science, NIST/University of Maryland, College Park, MD, 20742, USA}
 
  \author{ Alexey V. Gorshkov}
 \affiliation{Joint Quantum Institute and Joint Center for Quantum Information and Computer Science, NIST/University of Maryland, College Park, MD, 20742, USA}
 
\date{\today}
\begin{abstract}

We propose a realization of mesonic and baryonic quasiparticle excitations in Rydberg atom arrays with programmable interactions. 
Recent experiments have shown that such systems possess a $\mathbb{Z}_3$-ordered crystalline phase whose low-energy quasiparticles are defects in the crystalline order.  
By engineering a $\mathbb{Z}_3$-translational-symmetry breaking field on top of the Rydberg-blockaded Hamiltonian, we show that different types of defects experience confinement, and as a consequence form mesonic or baryonic quasiparticle excitations.
We illustrate the formation of these quasiparticles by studying a quantum chiral clock model related to  the Rydberg Hamiltonian. 
We then propose an experimental protocol involving out-of-equilibrium dynamics to directly probe the spectrum of the confined excitations.  We show that the confined quasiparticle spectrum can limit quantum information spreading  in this system.  This proposal is readily applicable to current Rydberg experiments, and the method can be easily generalized to more complex confined excitations (e.g. `tetraquarks', `pentaquarks') in phases with $\mathbb{Z}_q$ order for $q>3$.

\end{abstract}
\maketitle

\pagenumbering{arabic}

 The development of controllable and coherent quantum simulators has the potential to provide new insights into a variety of many-body systems~\cite{Cirac2012}. 
Such simulators are ideal for studying phenomena such as non-equilibrium physics or scattering in quantum field theories which are difficult to simulate classically~\cite{Eisert15, Gogolin2016, Preskill2018a,Jordan2014,Jordan2018a}.
One class of quantum many-body systems that has been of interest recently is those exhibiting confinement.
Confinement is the phenomenon whereby the fundamental excitations of a system experience a potential which increases indefinitely with their separation, resulting in the formation of bound states in the low-energy spectrum \cite{kogutreview, McCoy1977, Calabrese2017}.
This mechanism plays an important role in quantum chromodynamics (QCD), where confinement due to gauge fluctuations explains the formation of mesons and baryons from quarks.
Although confinement between quarks is well-established, there are a number of difficulties in obtaining  quantitative estimates for physical observables~\cite{Brambilla2014}.

Recently, there have been theoretical \cite{Surace2019,Banuls2019,Weimer2010a,Notarnicola2019,Magnifico2019,Fangli2019} and experimental \cite{tan2019observation}  works on quantum simulators realizing confinement.
Such quantum simulators realize experimental control over isolated quantum systems at the single-atom level, which allows a great deal of sensitivity in both manipulation and detection ~\cite{bernien2017probing,Barredo2018,Bloch2012}.
To date, these systems only exhibit pairwise confinement of particle-antiparticle pairs into mesonic two-particle bound states \footnote{Baryonic bound states have been proposed in Ref.~\onlinecite{Rapp2007}, but this model does not exhibit confinement.}.
To make closer contact with the phenomenology of QCD, it would be advantageous to realize a model Hamiltonian whose spectrum contains more complex bound states.

In this work, we propose a quantum simulator scheme to  implement confined baryonic and mesonic excitations in Rydberg atom arrays.
The basis for our proposal involves the recent realization of crystalline states  which exhibit spontaneously broken $\mathbb{Z}_q$ symmetry, where the chain is populated by a Rydberg excitation every $q$ sites \cite{bernien2017probing,Keesling2019}. The low-energy excitations above these ground states are defects which lie between the degenerate ordered ground states, and these defects may be separated from each other to arbitrary distances.
We show that by adding a non-uniform on-site detuning  which breaks the $\mathbb{Z}_q$ symmetry, the different types of domain-wall defects observed in Ref.~\cite{Keesling2019} are bound together so that the low-energy excitations are instead composite objects such as mesons or baryons. 
We demonstrate that the masses of these confined quasiparticles have a clear signature in the out-of-equilibrium  dynamics of the Hamiltonian, and that the correlation spreading of the system is dramatically reduced in the confined phase. We also discuss in detail  the initial state preparation and measurement scheme for observing these confined quasiparticles. 

\begin{figure}
  \centering\includegraphics[width=0.5\textwidth,height=5.0cm]{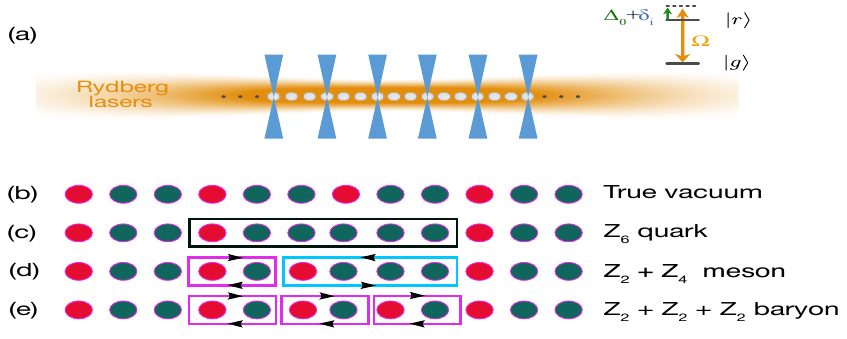}
  \caption{ (a) Implementation of symmetry breaking fields in the Rydberg array. The beams applied on every third atom create an additional detuning $\delta_i$ that is a nonzero constant only on these atoms. (b)-(e) Schematics of  mesonic and baryonic excitations formed in the Rydberg array. The red and green
  dots denote Rydberg  and ground states,  respectively.  With the symmetry breaking fields, different types of low-energy excitations can exist on top of (b) the true vacuum state, including (c) a $\mathbb{Z}_6$ quark excitation, (d) a mesonic excitation formed by $\mathbb{Z}_2 +\mathbb{Z}_4$ defects, and (e) a baryonic excitation formed by $\mathbb{Z}_2+\mathbb{Z}_2+\mathbb{Z}_2$ defects. 
  }
  \label{fig2}
\end{figure}

{\it The model.---} We study a one-dimensional array of Rydberg atoms  described by the  following Hamiltonian \cite{bernien2017probing}:

\begin{align}\label{eq:Rydberg_toy_confine}
H_{\rm Ryd}=&
\sum_{i} \frac{\Omega_i}{2} X_i - \sum_{i} \Delta_i n_i +\sum_{i<j} V_{|i-j|}n_i n_{j}.
\end{align}
Here, $\ket{r_i}$ ($\ket{g_i} $) denotes the Rydberg (ground) state for atom at site $i$ (Fig.\ \ref{fig2}), $X_i=\ket{g_i}\!\bra{r_i}+\ket{r_i}\!\bra{g_i} $, $n_i=\ket{r_i}\bra{r_i}$,  $\Omega_i$ and $\Delta_i$ are the Rabi frequencies and detunings respectively, and $V_{|i-j|}$ describes the interaction between atoms in the Rydberg state at sites $i$ and $j$. 
The interactions decay strongly with distance, with the scaling $V_{r} \propto 1/r^{6}$.
When both the Rabi frequencies and detunings are homogeneous in space, this Hamiltonian features ordered ground states where every $q$th site of the lattice is in the Rydberg state ($q\geq 2$) \cite{rader2019,FSS04,Samajdar2018,Chepiga2019}. Recent experiments have studied the quantum many-body dynamics governed by this Hamiltonian with homogeneous parameters \cite{bernien2017probing}, giving insight into phenomena such as quantum many-body scars \cite{turner2018,wenwei2019}, exotic quantum criticality \cite{Samajdar2018,Seth2018,rader2019}, and the quantum Kibble-Zurek mechanism  \cite{Keesling2019}.

{\it Relation to the $\mathbb{Z}_3$ chiral clock model.---} We focus on  the $\mathbb{Z}_3$ phase of the Rydberg Hamiltonian by specializing to the case where both $\Omega_i$ and $\Delta_i$ are much smaller than $V_1$.  
The low-energy quasiparticles above the ordered ground state [which is shown in Fig.~\ref{fig2}(b)]  are $\mathbb{Z}_2$ and $\mathbb{Z}_4$ domain walls between regions with $\mathbb{Z}_3$ order [see Fig.~\ref{fig2}(d-e)] \cite{Keesling2019}.
 We mention that these different types of domain walls (for the homogeneous case) have been directly observed in recent experiments \cite{Keesling2019}.
 At higher energies, one expects to furthermore get $\mathbb{Z}_q$ domain walls for $q>4$ [see, for example, Fig.~\ref{fig2}(c)].
 
\begin{figure}
  \centering\includegraphics[width=0.5\textwidth,height=5.0cm]{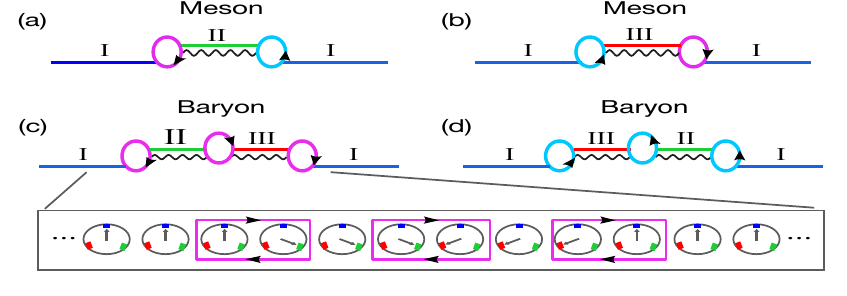}
  \caption{ Schematics of  mesonic and baryonic excitations formed by the three different vacua in the confined chiral clock model, Eq.~\eqref{ccmham}.  The different vacua are labelled by straight lines with different colors, and the clockwise (anti-clockwise) domain walls between two vacua are labelled by magenta (cyan) circles. The energy cost for creating domain walls scales linearly with their distances, as illustrated by the wavy line. For (c), we additionally give a schematic of the baryonic excitation in the clock basis.
  }
  \label{fig1}
\end{figure}

The system which displays similar physics with the ordered Rydberg system is the three-state quantum chiral clock model \cite{ostlund1981incommensurate,huse1982domain}
 \begin{equation}
\label{eq:Hamiltonian}
H_{\textsc{CCM0}} = -f \,\sum_{j} \tau_j^\dagger  - J \sum_{j} \sigma_j^\dagger \,\sigma_{j+1}\,\mathrm{e}^{-\mathrm{i}\,\theta} + \mbox{h.c.},
\end{equation}
where $\theta$ is a phase factor, and  the operators $\tau$ and $\sigma$ commute on different sites, with the matrix representation 
\renewcommand{\arraystretch}{0.85}
\begin{equation}
\sigma = 
\begin{pmatrix}
1 & 0 & 0 \\
0 & \omega & 0 \\
0 & 0 & \omega^2 \\
\end{pmatrix},
\qquad
\tau = 
\begin{pmatrix}
0 & 0 & 1 \\
1 & 0 & 0  \\
0 & 1 & 0 \\
\end{pmatrix},
\label{eq:basis}
\end{equation}
where $\omega = e^{2 \pi i/3}$.  The clock model obeys a $\mathbb{Z}_3$ symmetry generated by the operator $\mathcal{G}= \prod_i \tau_i$, and the  ground states (`vacua')  are three-fold degenerate in the ordered phase.  Consequently, the elementary excitations are  the clockwise or anti-clockwise domain walls  between any two different types of the three vacua [see Fig.~\ref{fig1}].  Because of the degeneracy, the domain walls may be separated to infinity, since moving a domain wall costs no energy.
For  $\theta = 0$, the two lowest defects have the same excitation energy, and Eq.~\eqref{eq:Hamiltonian} is the ordinary three-state quantum Potts model, while when $\theta \neq 0$, the two domain walls will have differing energies. 
Because the lowest-energy excitations have the same structure, the phase transitions in the clock and Rydberg models lie in the same universality class \cite{FSS04,Samajdar2018,Seth2018}.
Such clock models have been used to study low-energy confinement in a number of previous works \cite{McCoy1977,ZAM1989,delfino2008,Lepori2009,Rutkevich2015,lencses2015}, so we shall use intuition obtained from these works to understand how confinement can be engineered in the Rydberg arrays.

{\it Confinement via spatially periodic detunings.---} We now consider an additional spatially periodic detuning $\delta_i$ on top of the homogeneous $\Delta_0$ in the Rydberg Hamiltonian. 
Specifically, we analyze 
the case where there is an energy decrease of the Rydberg state for every third atom [see Fig.~\ref{fig2}(a)]. The periodic field can be realized in experiments by using  locally addressed lasers~\cite{Omran570}.  With this field, the three-fold degeneracy of the vacua and  the $\mathbb{Z}_3$ symmetry of Eq.~(\ref{eq:Hamiltonian}) [and Eq.~(\ref{eq:Rydberg_toy_confine})] are explicitly broken [see Fig.~\ref{fig1}]. We use \Romannum{1} to label the `true vacuum' which has lower energy than the other two degenerate `false vacua' \Romannum{2} and \Romannum{3} . The corresponding quantum clock Hamiltonian now includes an additional longitudinal field
\begin{equation}
H_{\textsc{CCM1}} = -f \,\sum_{j} \tau_j^\dagger  - J \sum_{j} \sigma_j^\dagger \,\sigma_{j+1}\,\mathrm{e}^{-\mathrm{i}\,\theta}   - h \sum_{j}  \sigma_j +\mbox{h.c.}
\label{ccmham}
\end{equation} 
 In contrast to the homogeneous case, with the longitudinal field, the domain-walls between true and false vacua cannot be separated to long distances due to a confining potential (an energy penalty) which scales linearly with the separation between defects. 
Consequently, the low-energy excitations of the Hamiltonian for large system sizes must be entirely made up of bound states of domain walls.
This is in close analogy to confinement in particle physics, where quarks cannot be directly observed in nature as two (three) of them are bound into mesons (baryons), due to similar confining potential scaling~ \cite{kogutreview, McCoy1977, Calabrese2017}.

Fig.~\ref{fig1} schematically shows the low-energy bound quasiparticle excitations formed by different vacua, which include both mesonic and baryonic bound states.  The mesonic  states are formed by two domain walls, including one clockwise (connecting I$\rightarrow$ II, II$\rightarrow $III, or III$\rightarrow $I) and one counterclockwise (connecting III$\rightarrow$ II, II$\rightarrow $I, or I$ \rightarrow $III) defect [Fig.~\ref{fig1}(a-b)]. On the other hand, the baryonic excitations are composed of either three clockwise domain walls or three anti-clockwise domain walls [see Fig.~\ref{fig1}(c-d)].  

A schematic plot of the low energy excitations on top of the true ground state for the Rydberg array is shown in Fig.~\ref{fig2}, where the picture is exact when $\Omega=0$.
The ordered ground state is mapped to the $\mathbb{Z}_3$-ordered crystalline state for the Rydberg chain [Fig.~\ref{fig2}(b)]. Further, the clockwise and anti-clockwise defects map to the $\mathbb{Z}_2$ and $\mathbb{Z}_4$  defects, respectively [Fig.~\ref{fig2}(d)]. Due to the same mechanism, the additional real-space periodic potential leads to a confining potential between the domain walls, which thus leads to bound states of the $\mathbb{Z}_2$ and $\mathbb{Z}_4$ defects.   It is clear that both the mesonic and baryonic excitations shown in Fig.~\ref{fig1} can be mapped to corresponding Rydberg configurations. 
We note that the Rydberg model of Eq.~\eqref{eq:Rydberg_toy_confine} will additionally allow higher-energy defect states such as ``$\mathbb{Z}_6$ quarks'' [Fig.~\ref{fig2}(c)], which have no analogue in the chiral clock model.

 Although we have focused on the $\mathbb{Z}_3$ case, the above can be formulated for any of the $\mathbb{Z}_q$ ordered states. 
For $q=2$, this corresponds to the well-studied confinement in the Ising model with a longitudinal field~\cite{Calabrese2017, Fangli2019}.
For $q>3$, one can additionally obtain more complicated `tetraquark' or `pentaquark' states, in which 4 or 5 domain walls, respectively, bind together.
We note that the spectrum of confined excitations for the Potts [$\theta = 0$ in Eq.\ (\ref{ccmham})] model has been explored in a number of theoretical works \cite{delfino2008,Lepori2009,Rutkevich2015,lencses2015}, but the general $\theta \neq 0$ case has not been explored yet. 
While our proposal to realize confinement does not come from a lattice gauge theory, the relation between confinement in spin models and gauge theories has a long history in the nonperturbative study of lattice gauge theories \cite{McCoy1977,elitzur1979,kogutreview,einhorn1980}.
This is particularly clear in one dimension, where the mechanism of confinement in gauge theories can be similarly described as binding together defects between degenerate vacua, although resulting from matter coupling to gauge field \cite{coleman1975,coleman1976}.
In fact, the $\mathbb{Z}_2$ case for our present system is dual to the bosonized massive Schwinger model \cite{multifreqSG,Shankar05}.

 \begin{figure}
  \centering\includegraphics[width=0.48\textwidth, height=9.5cm]{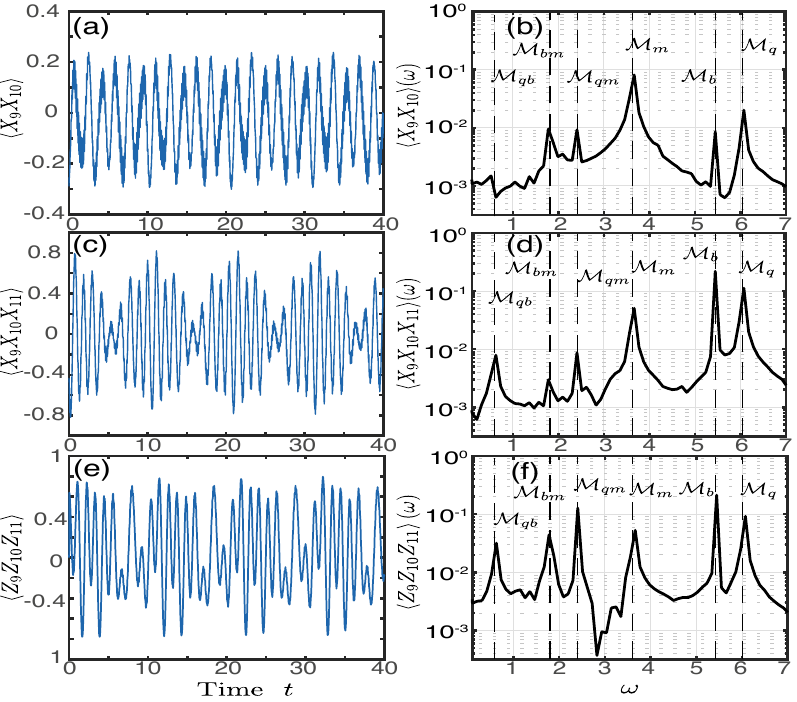}
  \caption{ (a-b) The initial state is chosen to maximize the total probability of $\ket{\Psi_v}$ and $\ket{\Psi_m}$ under the preparation protocol discussed in the text. (c-f) The initial state is chosen to maximize  total probability $\ket{\Psi_v}$ and $\ket{\Psi_b}$. Shown are the time-dependent expectation values and the associated Fourier spectra of (a-b) $\bra{\Psi(t)}X_9X_{10} \ket{\Psi(t)}$, (c-d) $\bra{\Psi(t)}X_9X_{10}X_{11} \ket{\Psi(t)}$, and (e-f) $\bra{\Psi(t)} U^\dagger Z_9Z_{10}Z_{11} U \ket{\Psi(t)}$, where $U$ is a $\pi/2$-pulse applied to the middle three atoms.  The dotted lines denote the energy of the meson ($\mathcal{M}_m$), baryon ($\mathcal{M}_b$), quark ($\mathcal{M}_q$) and their energy differences ($\mathcal{M}_{bm}$, $\mathcal{M}_{qb}$, $\mathcal{M}_{qm}$) obtained from exact diagonalization.  Parameters: $L=18$, $\Omega_i$ is chosen to be homogeneous $\Omega= 1;   \Delta_i=\Delta_0 +\delta_i,$ where $\Delta_0=4$ and $\delta_i= 2 $ for $(i \mod 3) =1$ and 0 otherwise; $ V_1=164.17$, corresponding to Rydberg radius being 2.34~\cite{Keesling2019}.}
  \label{fig3}
\end{figure}

 {\it Detecting quasiparticle masses.---}We   consider using quench dynamics to probe `mesonic' and `baryonic' masses in  Rydberg experiments.  One can in principle prepare the product state of Fig.~\ref{fig2}(b), and the subsequent quench dynamics under Eq.~(\ref{eq:Rydberg_toy_confine}) would be set by the excitation energies of the bound quasiparticles~\cite{Calabrese2017}. However, since we work in the  regime where $\Omega$ is much smaller than $\Delta$ and $V_2$, the excitation probability of bound quasiparticles can be  low as it intrinsically involves high-order processes \footnote{The transition from the vacuum state to the mesonic (baryonic) excitation involves a second (third) order process.}.

We instead choose the initial states to have sizable overlap with both the ground state and localized excited states.  Since the lowest-order mesonic and baryonic excitations  involve flipping  the states of only three atoms  (in the $\Omega_i=0$ limit),  we consider the set of initial states involving a large superposition of the true vacuum state $\ket{\Psi_{v}}= \ket{...rggrggrggrgg...}$ and the target mesonic $\ket{\Psi_{m}}= \ket{...rggrgrgggrgg...}$, or involving the true vacuum state and baryonic state $\ket{\Psi_{b}}= \ket{...rggrgrgrgrgg...}$ (see Fig.~\ref{fig2})\footnote{The `...' denotes repeated structure of the ordered phase with rgg}.
The intuition for choosing such initial states comes from the $\Omega_i=0$ limit, where the dynamics with such initial states involves oscillations between the `true vacuum' state and the localized mesonic and/or baryonic states displayed in Fig.~\ref{fig2}. 
We will show that the real-time dynamics involving these states can indeed resolve the many-body bound excitations for general $\Omega_i \neq 0$  [Fig.~\ref{fig3}].

We  first choose the time-dependent observables to be $X_9X_{10}X_{11}$ and $X_9X_{10}$ (chain length $L=18$), 
 as they have non-vanishing matrix elements between the vacuum state and the  baryonic  and mesonic state, respectively. 
 Fig.~\ref{fig3} shows the numerical results for the time-dependent expectation value of observables  and their Fourier transform. Note that the initial state is chosen to maximize the total probability of $\ket{\Psi_v}$ and $\ket{\Psi_m}$ [for Fig.~\ref{fig3}(a-b)], or the total probability of $\ket{\Psi_v}$ and $\ket{\Psi_
b}$  [for Fig.\ \ref{fig3}(c-d)], under a specific preparation protocol (discussed later).
  These states are evolved under the fully long-range Hamiltonian Eq.~\eqref{eq:Rydberg_toy_confine} with non-zero $\delta_i$, and other parameters are chosen such that the ground state is in the $\mathbb{Z}_3$-ordered phase for $\delta_i = 0$. As Figs.~\ref{fig3}(a) and (c) show,  the observables exhibit  clear  periodic oscillations. Their  Fourier spectra  [Fig.~\ref{fig3}(b) and (d)] agree perfectly with the masses of the `$\mathbb{Z}_2+\mathbb{Z}_4$ meson' ($\mathcal{M}_m$), the `$\mathbb{Z}_2+\mathbb{Z}_2+\mathbb{Z}_2$ baryon' ($\mathcal{M}_b$), the `$\mathbb{Z}_6$ quark' ($\mathcal{M}_q$), and their energy differences ($\mathcal{M}_{bm}$, $\mathcal{M}_{qb}$ and $\mathcal{M}_{qm}$). We mention that the highest Fourier peaks of Fig.~\ref{fig3} (b) and (d) agree with the mesonic and baryonic masses respectively, as the particular initial states have large components of the target excited states.

We note that in the current experiments only measurement in the $Z$-basis is possible.
In order to access the  masses given by $X$-observables, one can use a $\pi/2$ pulse (which we denote as $U$) rotating the states of the middle three atoms  before the measurement (subject to Rydberg constraint as discussed in detail later). 
The observable $\bra{\Psi(t)} U^\dagger Z_9Z_{10}Z_{11} U \ket{\Psi(t)}$ after the rotation also oscillates with a large contrast [see Figs.~\ref{fig3}(e)], and its Fourier spectrum accurately determines the masses of the baryonic, mesonic and the higher-energy quark excitations [Figs.~\ref{fig3}(f)]

\begin{figure}
  \centering\includegraphics[width=0.5\textwidth,height=3.5cm]{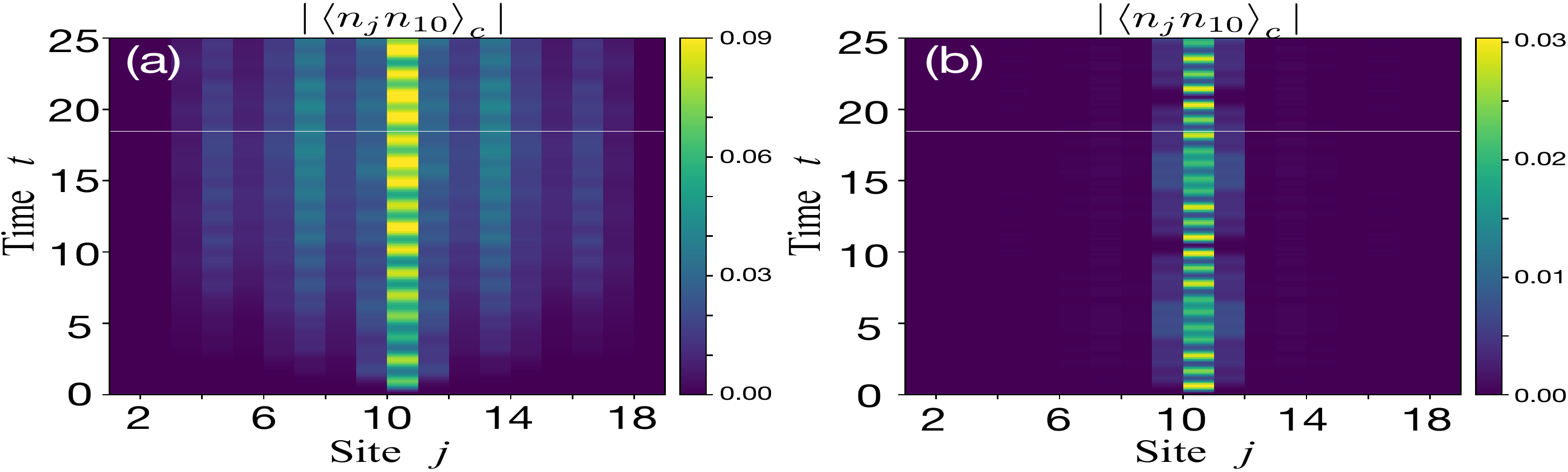}
  \caption{ Correlation spreading after a quantum quench from the $\mathbb{Z}_3$-ordered state $\ket{\Psi_v}$. (a) shows the time-dependent $ |\left\langle n_jn_{10} \right\rangle_c|$ without the additional periodic spatially detuning, i.e. $\delta_i=0$. (b) shows the same quantity with $\delta_i= 2$ for $(i \mod  3)=1$ and $0$ otherwise. All other parameters are the same for (a) and (b): $\Omega= 1,  \Delta_0=4,  V_1=164.17, L=19$.   }
  \label{fig4}
\end{figure}

{\it Quantum information spreading.---}The confined quasiparticles at low energy can have a dramatic effect on the correlation spreading in the system. For instance, confinement can strongly suppress the spreading of correlations and lead to slow thermalization \citep{Calabrese2017,zhicheng2019}. Here we focus on the time-dependent connected correlation function $ |\left\langle n_jn_k \right\rangle_c|= |\left\langle n_jn_k \right\rangle- \left\langle n_j\right\rangle \left\langle n_k\right\rangle|$ to study the quantum information spreading after a quantum quench from an initial $\mathbb{Z}_3$-ordered product state.  Fig.\ \ref{fig4}(a) shows the case of a homogeneous post-quench Hamiltonian with  $\delta_i=0$. As one can clearly see, correlations spread out fast across the 1D chain, leading to a light-cone structure \footnote{The light-cone shows a  $\mathbb{Z}_3$-periodic sub-structure due to the blockade physics.}. In contrast, with a periodic detuning field added to the Hamiltonian, the low-energy excitations are bound quasiparticles (mesons and baryons). In this case, the correlation spreads much slower than in the deconfined case [see Fig.~\ref{fig4}(b)]. We emphasize that such observables can also be directly measured in experiments \cite{bernien2017probing,Keesling2019}.

{\it Experimental preparation and detection.---}
To experimentally prepare the above sets of initial states, one can first prepare  the ordered product state $\ket{\Psi_{v}}= \ket{rggrggrggrggrggrgg}$ \cite{bernien2017probing,Keesling2019}, and rotate the 10th atom from the Rydberg to the ground state (via single-atom addressability), which yields $\ket{\Psi_{q}}= \ket{rggrggrgggggrggrgg}$ \footnote{The local rotation fields can be also engineered by applying a strong light shift to the selected target atoms, and shining an additional resonant Rydberg laser beam across the whole chain.}. 
After this,  a Rabi laser is shined only on the  middle three atoms, with a Rabi frequency $\Omega_0$ satisfying $V_2 \ll \Omega_0 \ll V_1$. 
To prepare the initial state for Fig.\ \ref{fig3}(a-b),  we apply the Rabi pulse for time $1.4 \pi/\Omega_0$.  
On the other hand, for the initial states in Fig.\ \ref{fig3}(c)-(f), we choose time $3.6 \pi/\Omega_0$. 
During the preparation, the parameters $\Delta$ and interactions $V_r$ are the same as in the post-quenched Hamiltonian [see Fig.~\ref{fig3}], while $\Omega_0=25$ for the three atoms, and all other atoms seeing vanishing Rabi frequencies.  We have checked that the probabilities for the blockaded states ($grr, rrg, rrr$) for the three middle atoms are on the order of  $10^{-3}$.

To measure the dynamics shown in Fig.~\ref{fig3}(e-f), after state preparation, we evolve the system under Hamiltonian (\ref{eq:Rydberg_toy_confine}), rotate the middle three atoms back using a $\pi/2$ pulse [i.e.~apply $\Omega_0$ for time $\pi/(2 \Omega_0)$], and finally measure $Z_9 Z_{10} Z_{11}$. During the $\pi/2$ pulse, all the parameters (except for the pulse time) are the same as for the preparation step.

 {\it Conclusions and outlook.---}Our proposal shows that Rydberg arrays are 
 a natural platform to study exotic confined excitations not only for the mesonic case, but also for baryonic quasiparticles. These confined excitations are analogous  
 to the more complicated bound states seen in high energy physics. 
 Although we focus on the regime where the Rydberg system is described by an effective clock model, we expect Eq.~\eqref{eq:Rydberg_toy_confine} to exhibit confinement wherever  the homogeneous Rydberg  model is in the $\mathbb{Z}_3$ ordered phase. Away from the clock limit, one needs to consider more general $\mathbb{Z}_q$ defect states where $q>4$, and this will lead to more complicated `hadronic' excitations. Correspondingly, this analysis can be generalised  to the other $\mathbb{Z}_q$-ordered phases of the Rydberg system, which will generally lead to a host of more complicated confined quasiparticles (e.g. `tetraquarks' and `pentaquarks').
 These more complex states would require much larger system sizes which would no longer be amenable to the numerical methods used here, but can be achieved in quantum simulators.
 Quantum simulators can additionally access dynamical phenomena such as string breaking and inelastic scattering which are intractable using classical methods \cite{Preskill2018a}.
 It would also be interesting to consider confinement scenarios in higher dimensions, where Rydberg systems feature more complicated phases of crystalline order \cite{Samajdar2020}. In particular, the symmetry-breaking patterns in two dimensions allow for both one-dimensional domain wall excitations as well as point-like ``monopole'' excitations, which is similar to the distinct excitations in higher-dimensional gauge theories \cite{einhorn1980}.
 
\begin{acknowledgments} We thank Zohreh  Davoudi, Sepehr Ebadi, Markus Heyl, Alexander Keesling, Mikhail Lukin, Ahmed Omran, and Roberto Verdel for insightful discussions. 
PB thanks ITAMP for hospitality and support via the Visitor Program.
This work was supported by AFOSR, AFOSR MURI, DoE BES Materials and Chemical Sciences Research for Quantum Information Science program (award No.~DE-SC0019449), DoE ASCR Quantum Testbed Pathfinder program (award No.~DE-SC0019040), DoE ASCR Accelerated Research in Quantum Computing program (award No.~DE-SC0020312), NSF PFCQC program, ARO MURI, ARL CDQI, and NSF PFC at JQI.
 
\end{acknowledgments}

\bibliography{confinement}

\begin{thebibliography}{50}%
\makeatletter
\providecommand \@ifxundefined [1]{%
 \@ifx{#1\undefined}
}%
\providecommand \@ifnum [1]{%
 \ifnum #1\expandafter \@firstoftwo
 \else \expandafter \@secondoftwo
 \fi
}%
\providecommand \@ifx [1]{%
 \ifx #1\expandafter \@firstoftwo
 \else \expandafter \@secondoftwo
 \fi
}%
\providecommand \natexlab [1]{#1}%
\providecommand \enquote  [1]{``#1''}%
\providecommand \bibnamefont  [1]{#1}%
\providecommand \bibfnamefont [1]{#1}%
\providecommand \citenamefont [1]{#1}%
\providecommand \href@noop [0]{\@secondoftwo}%
\providecommand \href [0]{\begingroup \@sanitize@url \@href}%
\providecommand \@href[1]{\@@startlink{#1}\@@href}%
\providecommand \@@href[1]{\endgroup#1\@@endlink}%
\providecommand \@sanitize@url [0]{\catcode `\\12\catcode `\$12\catcode
  `\&12\catcode `\#12\catcode `\^12\catcode `\_12\catcode `\%12\relax}%
\providecommand \@@startlink[1]{}%
\providecommand \@@endlink[0]{}%
\providecommand \url  [0]{\begingroup\@sanitize@url \@url }%
\providecommand \@url [1]{\endgroup\@href {#1}{\urlprefix }}%
\providecommand \urlprefix  [0]{URL }%
\providecommand \Eprint [0]{\href }%
\providecommand \doibase [0]{http://dx.doi.org/}%
\providecommand \selectlanguage [0]{\@gobble}%
\providecommand \bibinfo  [0]{\@secondoftwo}%
\providecommand \bibfield  [0]{\@secondoftwo}%
\providecommand \translation [1]{[#1]}%
\providecommand \BibitemOpen [0]{}%
\providecommand \bibitemStop [0]{}%
\providecommand \bibitemNoStop [0]{.\EOS\space}%
\providecommand \EOS [0]{\spacefactor3000\relax}%
\providecommand \BibitemShut  [1]{\csname bibitem#1\endcsname}%
\let\auto@bib@innerbib\@empty
\bibitem [{\citenamefont {Cirac}\ and\ \citenamefont
  {Zoller}(2012)}]{Cirac2012}%
  \BibitemOpen
  \bibfield  {author} {\bibinfo {author} {\bibfnamefont {J.~I.}\ \bibnamefont
  {Cirac}}\ and\ \bibinfo {author} {\bibfnamefont {P.}~\bibnamefont {Zoller}},\
  }\bibfield  {title} {\enquote {\bibinfo {title} {Goals and opportunities in
  quantum simulation},}\ }\href {https://doi.org/10.1038/nphys2275} {\bibfield
  {journal} {\bibinfo  {journal} {Nat. Phys.}\ }\textbf {\bibinfo {volume}
  {8}},\ \bibinfo {pages} {264} (\bibinfo {year} {2012})}\BibitemShut {NoStop}%
\bibitem [{\citenamefont {{Eisert}}\ \emph {et~al.}(2015)\citenamefont
  {{Eisert}}, \citenamefont {{Friesdorf}},\ and\ \citenamefont
  {{Gogolin}}}]{Eisert15}%
  \BibitemOpen
  \bibfield  {author} {\bibinfo {author} {\bibfnamefont {J.}~\bibnamefont
  {{Eisert}}}, \bibinfo {author} {\bibfnamefont {M.}~\bibnamefont
  {{Friesdorf}}}, \ and\ \bibinfo {author} {\bibfnamefont {C.}~\bibnamefont
  {{Gogolin}}},\ }\bibfield  {title} {\enquote {\bibinfo {title} {{Quantum
  many-body systems out of equilibrium}},}\ }\href {\doibase 10.1038/nphys3215}
  {\bibfield  {journal} {\bibinfo  {journal} {Nat. Phys.}\ }\textbf {\bibinfo
  {volume} {11}},\ \bibinfo {pages} {124} (\bibinfo {year} {2015})}\BibitemShut
  {NoStop}%
\bibitem [{\citenamefont {Gogolin}\ and\ \citenamefont
  {Eisert}(2016)}]{Gogolin2016}%
  \BibitemOpen
  \bibfield  {author} {\bibinfo {author} {\bibfnamefont {C.}~\bibnamefont
  {Gogolin}}\ and\ \bibinfo {author} {\bibfnamefont {J.}~\bibnamefont
  {Eisert}},\ }\bibfield  {title} {\enquote {\bibinfo {title} {Equilibration,
  thermalisation, and the emergence of statistical mechanics in closed quantum
  systems},}\ }\href {\doibase 10.1088/0034-4885/79/5/056001} {\bibfield
  {journal} {\bibinfo  {journal} {Rep. Prog. Phys.}\ }\textbf {\bibinfo
  {volume} {79}},\ \bibinfo {pages} {056001} (\bibinfo {year}
  {2016})}\BibitemShut {NoStop}%
\bibitem [{\citenamefont {{Preskill}}(2018)}]{Preskill2018a}%
  \BibitemOpen
  \bibfield  {author} {\bibinfo {author} {\bibfnamefont {J.}~\bibnamefont
  {{Preskill}}},\ }\bibfield  {title} {\enquote {\bibinfo {title} {{Simulating
  quantum field theory with a quantum computer}},}\ }in\ \href
  {https://ui.adsabs.harvard.edu/abs/2018slft.confE..24P} {\emph {\bibinfo
  {booktitle} {The 36th Annual International Symposium on Lattice Field Theory.
  22-28 July}}}\ (\bibinfo {year} {2018})\ \Eprint
  {http://arxiv.org/abs/1811.10085} {arXiv:1811.10085 [hep-lat]} \BibitemShut
  {NoStop}%
\bibitem [{\citenamefont {{Jordan}}\ \emph {et~al.}(2014)\citenamefont
  {{Jordan}}, \citenamefont {{Lee}},\ and\ \citenamefont
  {{Preskill}}}]{Jordan2014}%
  \BibitemOpen
  \bibfield  {author} {\bibinfo {author} {\bibfnamefont {S.~P.}\ \bibnamefont
  {{Jordan}}}, \bibinfo {author} {\bibfnamefont {K.~S.~M.}\ \bibnamefont
  {{Lee}}}, \ and\ \bibinfo {author} {\bibfnamefont {J.}~\bibnamefont
  {{Preskill}}},\ }\bibfield  {title} {\enquote {\bibinfo {title} {{Quantum
  Computation of Scattering in Scalar Quantum Field Theories}},}\ }\href
  {https://dl.acm.org/doi/10.5555/2685155.2685163} {\bibfield  {journal}
  {\bibinfo  {journal} {Quantum Inf. Comput.}\ }\textbf {\bibinfo {volume}
  {2}},\ \bibinfo {pages} {1014} (\bibinfo {year} {2014})}\BibitemShut
  {NoStop}%
\bibitem [{\citenamefont {Jordan}\ \emph {et~al.}(2018)\citenamefont {Jordan},
  \citenamefont {Krovi}, \citenamefont {Lee},\ and\ \citenamefont
  {Preskill}}]{Jordan2018a}%
  \BibitemOpen
  \bibfield  {author} {\bibinfo {author} {\bibfnamefont {S.~P.}\ \bibnamefont
  {Jordan}}, \bibinfo {author} {\bibfnamefont {H.}~\bibnamefont {Krovi}},
  \bibinfo {author} {\bibfnamefont {K.~S.~M.}\ \bibnamefont {Lee}}, \ and\
  \bibinfo {author} {\bibfnamefont {J.}~\bibnamefont {Preskill}},\ }\bibfield
  {title} {\enquote {\bibinfo {title} {{BQP}-completeness of scattering in
  scalar quantum field theory},}\ }\href {\doibase 10.22331/q-2018-01-08-44}
  {\bibfield  {journal} {\bibinfo  {journal} {{Quantum}}\ }\textbf {\bibinfo
  {volume} {2}},\ \bibinfo {pages} {44} (\bibinfo {year} {2018})}\BibitemShut
  {NoStop}%
\bibitem [{\citenamefont {Kogut}(1979)}]{kogutreview}%
  \BibitemOpen
  \bibfield  {author} {\bibinfo {author} {\bibfnamefont {J.~B.}\ \bibnamefont
  {Kogut}},\ }\bibfield  {title} {\enquote {\bibinfo {title} {An introduction
  to lattice gauge theory and spin systems},}\ }\href {\doibase
  10.1103/RevModPhys.51.659} {\bibfield  {journal} {\bibinfo  {journal} {Rev.
  Mod. Phys.}\ }\textbf {\bibinfo {volume} {51}},\ \bibinfo {pages} {659}
  (\bibinfo {year} {1979})}\BibitemShut {NoStop}%
\bibitem [{\citenamefont {McCoy}\ and\ \citenamefont {Wu}(1977)}]{McCoy1977}%
  \BibitemOpen
  \bibfield  {author} {\bibinfo {author} {\bibfnamefont {B.~M.}\ \bibnamefont
  {McCoy}}\ and\ \bibinfo {author} {\bibfnamefont {T.~T.}\ \bibnamefont {Wu}},\
  }\bibfield  {title} {\enquote {\bibinfo {title} {Speculations on quark
  observation},}\ }\href {\doibase
  https://doi.org/10.1016/0370-2693(77)90706-7} {\bibfield  {journal} {\bibinfo
   {journal} {Phys. Lett. B}\ }\textbf {\bibinfo {volume} {72}},\ \bibinfo
  {pages} {219} (\bibinfo {year} {1977})}\BibitemShut {NoStop}%
\bibitem [{\citenamefont {{Kormos}}\ \emph {et~al.}(2017)\citenamefont
  {{Kormos}}, \citenamefont {{Collura}}, \citenamefont {{Tak{\'a}cs}},\ and\
  \citenamefont {{Calabrese}}}]{Calabrese2017}%
  \BibitemOpen
  \bibfield  {author} {\bibinfo {author} {\bibfnamefont {M.}~\bibnamefont
  {{Kormos}}}, \bibinfo {author} {\bibfnamefont {M.}~\bibnamefont {{Collura}}},
  \bibinfo {author} {\bibfnamefont {G.}~\bibnamefont {{Tak{\'a}cs}}}, \ and\
  \bibinfo {author} {\bibfnamefont {P.}~\bibnamefont {{Calabrese}}},\
  }\bibfield  {title} {\enquote {\bibinfo {title} {{Real-time confinement
  following a quantum quench to a non-integrable model}},}\ }\href {\doibase
  10.1038/nphys3934} {\bibfield  {journal} {\bibinfo  {journal} {Nat. Phys.}\
  }\textbf {\bibinfo {volume} {13}},\ \bibinfo {pages} {246} (\bibinfo {year}
  {2017})}\BibitemShut {NoStop}%
\bibitem [{\citenamefont {Brambilla}\ \emph {et~al.}(2014)\citenamefont
  {Brambilla} \emph {et~al.}}]{Brambilla2014}%
  \BibitemOpen
  \bibfield  {author} {\bibinfo {author} {\bibfnamefont {N.}~\bibnamefont
  {Brambilla}} \emph {et~al.},\ }\bibfield  {title} {\enquote {\bibinfo {title}
  {{QCD and Strongly Coupled Gauge Theories: Challenges and Perspectives}},}\
  }\href {\doibase 10.1140/epjc/s10052-014-2981-5} {\bibfield  {journal}
  {\bibinfo  {journal} {Eur.\ Phys.\ J.\ C}\ }\textbf {\bibinfo {volume}
  {74}},\ \bibinfo {pages} {2981} (\bibinfo {year} {2014})}\BibitemShut
  {NoStop}%
\bibitem [{\citenamefont {Surace}\ \emph {et~al.}(2019)\citenamefont {Surace},
  \citenamefont {Russomanno}, \citenamefont {Dalmonte}, \citenamefont {Silva},
  \citenamefont {Fazio},\ and\ \citenamefont {Iemini}}]{Surace2019}%
  \BibitemOpen
  \bibfield  {author} {\bibinfo {author} {\bibfnamefont {F.~M.}\ \bibnamefont
  {Surace}}, \bibinfo {author} {\bibfnamefont {A.}~\bibnamefont {Russomanno}},
  \bibinfo {author} {\bibfnamefont {M.}~\bibnamefont {Dalmonte}}, \bibinfo
  {author} {\bibfnamefont {A.}~\bibnamefont {Silva}}, \bibinfo {author}
  {\bibfnamefont {R.}~\bibnamefont {Fazio}}, \ and\ \bibinfo {author}
  {\bibfnamefont {F.}~\bibnamefont {Iemini}},\ }\bibfield  {title} {\enquote
  {\bibinfo {title} {Floquet time crystals in clock models},}\ }\href {\doibase
  10.1103/PhysRevB.99.104303} {\bibfield  {journal} {\bibinfo  {journal} {Phys.
  Rev. B}\ }\textbf {\bibinfo {volume} {99}},\ \bibinfo {pages} {104303}
  (\bibinfo {year} {2019})}\BibitemShut {NoStop}%
\bibitem [{\citenamefont {{Ba{\~n}uls}}\ \emph {et~al.}()\citenamefont
  {{Ba{\~n}uls}}, \citenamefont {{Blatt}}, \citenamefont {{Catani}},
  \citenamefont {{Celi}}, \citenamefont {{Cirac}}, \citenamefont {{Dalmonte}},
  \citenamefont {{Fallani}}, \citenamefont {{Jansen}}, \citenamefont
  {{Lewenstein}}, \citenamefont {{Montangero}}, \citenamefont {{Muschik}},
  \citenamefont {{Reznik}}, \citenamefont {{Rico}}, \citenamefont
  {{Tagliacozzo}}, \citenamefont {{Van Acoleyen}}, \citenamefont
  {{Verstraete}}, \citenamefont {{Wiese}}, \citenamefont {{Wingate}},
  \citenamefont {{Zakrzewski}},\ and\ \citenamefont {{Zoller}}}]{Banuls2019}%
  \BibitemOpen
  \bibfield  {author} {\bibinfo {author} {\bibfnamefont {M.~C.}\ \bibnamefont
  {{Ba{\~n}uls}}}, \bibinfo {author} {\bibfnamefont {R.}~\bibnamefont
  {{Blatt}}}, \bibinfo {author} {\bibfnamefont {J.}~\bibnamefont {{Catani}}},
  \bibinfo {author} {\bibfnamefont {A.}~\bibnamefont {{Celi}}}, \bibinfo
  {author} {\bibfnamefont {J.~I.}\ \bibnamefont {{Cirac}}}, \bibinfo {author}
  {\bibfnamefont {M.}~\bibnamefont {{Dalmonte}}}, \bibinfo {author}
  {\bibfnamefont {L.}~\bibnamefont {{Fallani}}}, \bibinfo {author}
  {\bibfnamefont {K.}~\bibnamefont {{Jansen}}}, \bibinfo {author}
  {\bibfnamefont {M.}~\bibnamefont {{Lewenstein}}}, \bibinfo {author}
  {\bibfnamefont {S.}~\bibnamefont {{Montangero}}}, \bibinfo {author}
  {\bibfnamefont {C.~A.}\ \bibnamefont {{Muschik}}}, \bibinfo {author}
  {\bibfnamefont {B.}~\bibnamefont {{Reznik}}}, \bibinfo {author}
  {\bibfnamefont {E.}~\bibnamefont {{Rico}}}, \bibinfo {author} {\bibfnamefont
  {L.}~\bibnamefont {{Tagliacozzo}}}, \bibinfo {author} {\bibfnamefont
  {K.}~\bibnamefont {{Van Acoleyen}}}, \bibinfo {author} {\bibfnamefont
  {F.}~\bibnamefont {{Verstraete}}}, \bibinfo {author} {\bibfnamefont {U.~J.}\
  \bibnamefont {{Wiese}}}, \bibinfo {author} {\bibfnamefont {M.}~\bibnamefont
  {{Wingate}}}, \bibinfo {author} {\bibfnamefont {J.}~\bibnamefont
  {{Zakrzewski}}}, \ and\ \bibinfo {author} {\bibfnamefont {P.}~\bibnamefont
  {{Zoller}}},\ }\bibfield  {title} {\enquote {\bibinfo {title} {{Simulating
  Lattice Gauge Theories within Quantum Technologies}},}\ }\href
  {https://ui.adsabs.harvard.edu/abs/2019arXiv191100003B} {\ }\Eprint
  {http://arxiv.org/abs/1911.00003} {arXiv:1911.00003} \BibitemShut {NoStop}%
\bibitem [{\citenamefont {{Weimer}}\ \emph {et~al.}(2010)\citenamefont
  {{Weimer}}, \citenamefont {{M{\"u}ller}}, \citenamefont {{Lesanovsky}},
  \citenamefont {{Zoller}},\ and\ \citenamefont {{B{\"u}chler}}}]{Weimer2010a}%
  \BibitemOpen
  \bibfield  {author} {\bibinfo {author} {\bibfnamefont {H.}~\bibnamefont
  {{Weimer}}}, \bibinfo {author} {\bibfnamefont {M.}~\bibnamefont
  {{M{\"u}ller}}}, \bibinfo {author} {\bibfnamefont {I.}~\bibnamefont
  {{Lesanovsky}}}, \bibinfo {author} {\bibfnamefont {P.}~\bibnamefont
  {{Zoller}}}, \ and\ \bibinfo {author} {\bibfnamefont {H.~P.}\ \bibnamefont
  {{B{\"u}chler}}},\ }\bibfield  {title} {\enquote {\bibinfo {title} {{A
  Rydberg quantum simulator}},}\ }\href {\doibase 10.1038/nphys1614} {\bibfield
   {journal} {\bibinfo  {journal} {Nat. Phys.}\ }\textbf {\bibinfo {volume}
  {6}},\ \bibinfo {pages} {382} (\bibinfo {year} {2010})}\BibitemShut {NoStop}%
\bibitem [{\citenamefont {Notarnicola}\ \emph {et~al.}(2020)\citenamefont
  {Notarnicola}, \citenamefont {Collura},\ and\ \citenamefont
  {Montangero}}]{Notarnicola2019}%
  \BibitemOpen
  \bibfield  {author} {\bibinfo {author} {\bibfnamefont {S.}~\bibnamefont
  {Notarnicola}}, \bibinfo {author} {\bibfnamefont {M.}~\bibnamefont
  {Collura}}, \ and\ \bibinfo {author} {\bibfnamefont {S.}~\bibnamefont
  {Montangero}},\ }\bibfield  {title} {\enquote {\bibinfo {title}
  {Real-time-dynamics quantum simulation of $(1+1)\text{-dimensional}$ lattice
  qed with rydberg atoms},}\ }\href {\doibase 10.1103/PhysRevResearch.2.013288}
  {\bibfield  {journal} {\bibinfo  {journal} {Phys. Rev. Research}\ }\textbf
  {\bibinfo {volume} {2}},\ \bibinfo {pages} {013288} (\bibinfo {year}
  {2020})}\BibitemShut {NoStop}%
\bibitem [{\citenamefont {{Magnifico}}\ \emph {et~al.}()\citenamefont
  {{Magnifico}}, \citenamefont {{Dalmonte}}, \citenamefont {{Facchi}},
  \citenamefont {{Pascazio}}, \citenamefont {{Pepe}},\ and\ \citenamefont
  {{Ercolessi}}}]{Magnifico2019}%
  \BibitemOpen
  \bibfield  {author} {\bibinfo {author} {\bibfnamefont {G.}~\bibnamefont
  {{Magnifico}}}, \bibinfo {author} {\bibfnamefont {M.}~\bibnamefont
  {{Dalmonte}}}, \bibinfo {author} {\bibfnamefont {P.}~\bibnamefont
  {{Facchi}}}, \bibinfo {author} {\bibfnamefont {S.}~\bibnamefont
  {{Pascazio}}}, \bibinfo {author} {\bibfnamefont {F.~V.}\ \bibnamefont
  {{Pepe}}}, \ and\ \bibinfo {author} {\bibfnamefont {E.}~\bibnamefont
  {{Ercolessi}}},\ }\bibfield  {title} {\enquote {\bibinfo {title} {{Real Time
  Dynamics and Confinement in the $\mathbb{Z}_{n}$ Schwinger-Weyl lattice model
  for 1+1 QED}},}\ }\href {\doibase
  https://ui.adsabs.harvard.edu/abs/2019arXiv190904821M} {\ ,\ \bibinfo {pages}
  {arXiv:1909.04821}}\BibitemShut {NoStop}%
\bibitem [{\citenamefont {Liu}\ \emph {et~al.}(2019)\citenamefont {Liu},
  \citenamefont {Lundgren}, \citenamefont {Titum}, \citenamefont {Pagano},
  \citenamefont {Zhang}, \citenamefont {Monroe},\ and\ \citenamefont
  {Gorshkov}}]{Fangli2019}%
  \BibitemOpen
  \bibfield  {author} {\bibinfo {author} {\bibfnamefont {F.}~\bibnamefont
  {Liu}}, \bibinfo {author} {\bibfnamefont {R.}~\bibnamefont {Lundgren}},
  \bibinfo {author} {\bibfnamefont {P.}~\bibnamefont {Titum}}, \bibinfo
  {author} {\bibfnamefont {G.}~\bibnamefont {Pagano}}, \bibinfo {author}
  {\bibfnamefont {J.}~\bibnamefont {Zhang}}, \bibinfo {author} {\bibfnamefont
  {C.}~\bibnamefont {Monroe}}, \ and\ \bibinfo {author} {\bibfnamefont {A.~V.}\
  \bibnamefont {Gorshkov}},\ }\bibfield  {title} {\enquote {\bibinfo {title}
  {Confined quasiparticle dynamics in long-range interacting quantum spin
  chains},}\ }\href {\doibase 10.1103/PhysRevLett.122.150601} {\bibfield
  {journal} {\bibinfo  {journal} {Phys. Rev. Lett.}\ }\textbf {\bibinfo
  {volume} {122}},\ \bibinfo {pages} {150601} (\bibinfo {year}
  {2019})}\BibitemShut {NoStop}%
\bibitem [{\citenamefont {{Tan}}\ \emph {et~al.}()\citenamefont {{Tan}},
  \citenamefont {{Becker}}, \citenamefont {{Liu}}, \citenamefont {{Pagano}},
  \citenamefont {{Collins}}, \citenamefont {{De}}, \citenamefont {{Feng}},
  \citenamefont {{Kaplan}}, \citenamefont {{Kyprianidis}}, \citenamefont
  {{Lundgren}}, \citenamefont {{Morong}}, \citenamefont {{Whitsitt}},
  \citenamefont {{Gorshkov}},\ and\ \citenamefont
  {{Monroe}}}]{tan2019observation}%
  \BibitemOpen
  \bibfield  {author} {\bibinfo {author} {\bibfnamefont {W.~L.}\ \bibnamefont
  {{Tan}}}, \bibinfo {author} {\bibfnamefont {P.}~\bibnamefont {{Becker}}},
  \bibinfo {author} {\bibfnamefont {F.}~\bibnamefont {{Liu}}}, \bibinfo
  {author} {\bibfnamefont {G.}~\bibnamefont {{Pagano}}}, \bibinfo {author}
  {\bibfnamefont {K.~S.}\ \bibnamefont {{Collins}}}, \bibinfo {author}
  {\bibfnamefont {A.}~\bibnamefont {{De}}}, \bibinfo {author} {\bibfnamefont
  {L.}~\bibnamefont {{Feng}}}, \bibinfo {author} {\bibfnamefont {H.~B.}\
  \bibnamefont {{Kaplan}}}, \bibinfo {author} {\bibfnamefont {A.}~\bibnamefont
  {{Kyprianidis}}}, \bibinfo {author} {\bibfnamefont {R.}~\bibnamefont
  {{Lundgren}}}, \bibinfo {author} {\bibfnamefont {W.}~\bibnamefont
  {{Morong}}}, \bibinfo {author} {\bibfnamefont {S.}~\bibnamefont
  {{Whitsitt}}}, \bibinfo {author} {\bibfnamefont {A.~V.}\ \bibnamefont
  {{Gorshkov}}}, \ and\ \bibinfo {author} {\bibfnamefont {C.}~\bibnamefont
  {{Monroe}}},\ }\bibfield  {title} {\enquote {\bibinfo {title} {{Observation
  of Domain Wall Confinement and Dynamics in a Quantum Simulator}},}\ }\href
  {https://arxiv.org/abs/1912.11117} {\ }\Eprint
  {http://arxiv.org/abs/1912.11117} {arXiv:1912.11117} \BibitemShut {NoStop}%
\bibitem [{\citenamefont {Bernien}\ \emph {et~al.}(2017)\citenamefont
  {Bernien}, \citenamefont {Schwartz}, \citenamefont {Keesling}, \citenamefont
  {Levine}, \citenamefont {Omran}, \citenamefont {Pichler}, \citenamefont
  {Choi}, \citenamefont {Zibrov}, \citenamefont {Endres}, \citenamefont
  {Greiner}, \citenamefont {Vuleti{\'c}},\ and\ \citenamefont
  {Lukin}}]{bernien2017probing}%
  \BibitemOpen
  \bibfield  {author} {\bibinfo {author} {\bibfnamefont {H.}~\bibnamefont
  {Bernien}}, \bibinfo {author} {\bibfnamefont {S.}~\bibnamefont {Schwartz}},
  \bibinfo {author} {\bibfnamefont {A.}~\bibnamefont {Keesling}}, \bibinfo
  {author} {\bibfnamefont {H.}~\bibnamefont {Levine}}, \bibinfo {author}
  {\bibfnamefont {A.}~\bibnamefont {Omran}}, \bibinfo {author} {\bibfnamefont
  {H.}~\bibnamefont {Pichler}}, \bibinfo {author} {\bibfnamefont
  {S.}~\bibnamefont {Choi}}, \bibinfo {author} {\bibfnamefont {A.~S.}\
  \bibnamefont {Zibrov}}, \bibinfo {author} {\bibfnamefont {M.}~\bibnamefont
  {Endres}}, \bibinfo {author} {\bibfnamefont {M.}~\bibnamefont {Greiner}},
  \bibinfo {author} {\bibfnamefont {V.}~\bibnamefont {Vuleti{\'c}}}, \ and\
  \bibinfo {author} {\bibfnamefont {M.~D.}\ \bibnamefont {Lukin}},\ }\bibfield
  {title} {\enquote {\bibinfo {title} {Probing many-body dynamics on a 51-atom
  quantum simulator},}\ }\href {\doibase 10.1038/nature24622} {\bibfield
  {journal} {\bibinfo  {journal} {Nature}\ }\textbf {\bibinfo {volume} {551}},\
  \bibinfo {pages} {579} (\bibinfo {year} {2017})}\BibitemShut {NoStop}%
\bibitem [{\citenamefont {Barredo}\ \emph {et~al.}(2018)\citenamefont
  {Barredo}, \citenamefont {Lienhard}, \citenamefont {de~L{\'{e}}s{\'{e}}leuc},
  \citenamefont {Lahaye},\ and\ \citenamefont {Browaeys}}]{Barredo2018}%
  \BibitemOpen
  \bibfield  {author} {\bibinfo {author} {\bibfnamefont {D.}~\bibnamefont
  {Barredo}}, \bibinfo {author} {\bibfnamefont {V.}~\bibnamefont {Lienhard}},
  \bibinfo {author} {\bibfnamefont {S.}~\bibnamefont
  {de~L{\'{e}}s{\'{e}}leuc}}, \bibinfo {author} {\bibfnamefont
  {T.}~\bibnamefont {Lahaye}}, \ and\ \bibinfo {author} {\bibfnamefont
  {A.}~\bibnamefont {Browaeys}},\ }\bibfield  {title} {\enquote {\bibinfo
  {title} {{Synthetic three-dimensional atomic structures assembled atom by
  atom}},}\ }\href {\doibase 10.1038/s41586-018-0450-2} {\bibfield  {journal}
  {\bibinfo  {journal} {Nature}\ }\textbf {\bibinfo {volume} {561}},\ \bibinfo
  {pages} {79} (\bibinfo {year} {2018})}\BibitemShut {NoStop}%
\bibitem [{\citenamefont {Bloch}\ \emph {et~al.}(2012)\citenamefont {Bloch},
  \citenamefont {Dalibard},\ and\ \citenamefont
  {Nascimb{\`{e}}ne}}]{Bloch2012}%
  \BibitemOpen
  \bibfield  {author} {\bibinfo {author} {\bibfnamefont {I.}~\bibnamefont
  {Bloch}}, \bibinfo {author} {\bibfnamefont {J.}~\bibnamefont {Dalibard}}, \
  and\ \bibinfo {author} {\bibfnamefont {S.}~\bibnamefont {Nascimb{\`{e}}ne}},\
  }\bibfield  {title} {\enquote {\bibinfo {title} {{Quantum simulations with
  ultracold quantum gases}},}\ }\href {\doibase 10.1038/nphys2259} {\bibfield
  {journal} {\bibinfo  {journal} {Nat. Phys.}\ }\textbf {\bibinfo {volume}
  {8}},\ \bibinfo {pages} {267} (\bibinfo {year} {2012})}\BibitemShut {NoStop}%
\bibitem [{Note1()}]{Note1}%
  \BibitemOpen
  \bibinfo {note} {Baryonic bound states have been proposed in Ref.~\protect
  \rev@citealp {Rapp2007}, but this model does not exhibit
  confinement.}\BibitemShut {Stop}%
\bibitem [{\citenamefont {{Keesling}}\ \emph {et~al.}(2019)\citenamefont
  {{Keesling}}, \citenamefont {{Omran}}, \citenamefont {{Levine}},
  \citenamefont {{Bernien}}, \citenamefont {{Pichler}}, \citenamefont {{Choi}},
  \citenamefont {{Samajdar}}, \citenamefont {{Schwartz}}, \citenamefont
  {{Silvi}}, \citenamefont {{Sachdev}}, \citenamefont {{Zoller}}, \citenamefont
  {{Endres}}, \citenamefont {{Greiner}}, \citenamefont {{Vuleti{\'c}}},
  \citenamefont {{}},\ and\ \citenamefont {{Lukin}}}]{Keesling2019}%
  \BibitemOpen
  \bibfield  {author} {\bibinfo {author} {\bibfnamefont {A.}~\bibnamefont
  {{Keesling}}}, \bibinfo {author} {\bibfnamefont {A.}~\bibnamefont {{Omran}}},
  \bibinfo {author} {\bibfnamefont {H.}~\bibnamefont {{Levine}}}, \bibinfo
  {author} {\bibfnamefont {H.}~\bibnamefont {{Bernien}}}, \bibinfo {author}
  {\bibfnamefont {H.}~\bibnamefont {{Pichler}}}, \bibinfo {author}
  {\bibfnamefont {S.}~\bibnamefont {{Choi}}}, \bibinfo {author} {\bibfnamefont
  {R.}~\bibnamefont {{Samajdar}}}, \bibinfo {author} {\bibfnamefont
  {S.}~\bibnamefont {{Schwartz}}}, \bibinfo {author} {\bibfnamefont
  {P.}~\bibnamefont {{Silvi}}}, \bibinfo {author} {\bibfnamefont
  {S.}~\bibnamefont {{Sachdev}}}, \bibinfo {author} {\bibfnamefont
  {P.}~\bibnamefont {{Zoller}}}, \bibinfo {author} {\bibfnamefont
  {M.}~\bibnamefont {{Endres}}}, \bibinfo {author} {\bibfnamefont
  {M.}~\bibnamefont {{Greiner}}}, \bibinfo {author} {\bibnamefont
  {{Vuleti{\'c}}}}, \bibinfo {author} {\bibfnamefont {V.}~\bibnamefont {{}}}, \
  and\ \bibinfo {author} {\bibfnamefont {M.~D.}\ \bibnamefont {{Lukin}}},\
  }\bibfield  {title} {\enquote {\bibinfo {title} {{Quantum Kibble-Zurek
  mechanism and critical dynamics on a programmable Rydberg simulator}},}\
  }\href {\doibase 10.1038/s41586-019-1070-1} {\bibfield  {journal} {\bibinfo
  {journal} {Nature}\ }\textbf {\bibinfo {volume} {568}},\ \bibinfo {pages}
  {207} (\bibinfo {year} {2019})}\BibitemShut {NoStop}%
\bibitem [{\citenamefont {{Rader}}\ and\ \citenamefont
  {{L{\"a}uchli}}()}]{rader2019}%
  \BibitemOpen
  \bibfield  {author} {\bibinfo {author} {\bibfnamefont {M.}~\bibnamefont
  {{Rader}}}\ and\ \bibinfo {author} {\bibfnamefont {A.~M.}\ \bibnamefont
  {{L{\"a}uchli}}},\ }\bibfield  {title} {\enquote {\bibinfo {title} {{Floating
  Phases in One-Dimensional Rydberg Ising Chains}},}\ }\href@noop {} {\
  }\Eprint {http://arxiv.org/abs/1908.02068} {arXiv:1908.02068} \BibitemShut
  {NoStop}%
\bibitem [{\citenamefont {Fendley}\ \emph {et~al.}(2004)\citenamefont
  {Fendley}, \citenamefont {Sengupta},\ and\ \citenamefont {Sachdev}}]{FSS04}%
  \BibitemOpen
  \bibfield  {author} {\bibinfo {author} {\bibfnamefont {P.}~\bibnamefont
  {Fendley}}, \bibinfo {author} {\bibfnamefont {K.}~\bibnamefont {Sengupta}}, \
  and\ \bibinfo {author} {\bibfnamefont {S.}~\bibnamefont {Sachdev}},\
  }\bibfield  {title} {\enquote {\bibinfo {title} {Competing density-wave
  orders in a one-dimensional hard-boson model},}\ }\href {\doibase
  10.1103/PhysRevB.69.075106} {\bibfield  {journal} {\bibinfo  {journal} {Phys.
  Rev. B}\ }\textbf {\bibinfo {volume} {69}},\ \bibinfo {pages} {075106}
  (\bibinfo {year} {2004})}\BibitemShut {NoStop}%
\bibitem [{\citenamefont {Samajdar}\ \emph {et~al.}(2018)\citenamefont
  {Samajdar}, \citenamefont {Choi}, \citenamefont {Pichler}, \citenamefont
  {Lukin},\ and\ \citenamefont {Sachdev}}]{Samajdar2018}%
  \BibitemOpen
  \bibfield  {author} {\bibinfo {author} {\bibfnamefont {R.}~\bibnamefont
  {Samajdar}}, \bibinfo {author} {\bibfnamefont {S.}~\bibnamefont {Choi}},
  \bibinfo {author} {\bibfnamefont {H.}~\bibnamefont {Pichler}}, \bibinfo
  {author} {\bibfnamefont {M.~D.}\ \bibnamefont {Lukin}}, \ and\ \bibinfo
  {author} {\bibfnamefont {S.}~\bibnamefont {Sachdev}},\ }\bibfield  {title}
  {\enquote {\bibinfo {title} {Numerical study of the chiral $\mathbb{Z}_{3}$
  quantum phase transition in one spatial dimension},}\ }\href {\doibase
  10.1103/PhysRevA.98.023614} {\bibfield  {journal} {\bibinfo  {journal} {Phys.
  Rev. A}\ }\textbf {\bibinfo {volume} {98}},\ \bibinfo {pages} {023614}
  (\bibinfo {year} {2018})}\BibitemShut {NoStop}%
\bibitem [{\citenamefont {Chepiga}\ and\ \citenamefont
  {Mila}(2019)}]{Chepiga2019}%
  \BibitemOpen
  \bibfield  {author} {\bibinfo {author} {\bibfnamefont {N.}~\bibnamefont
  {Chepiga}}\ and\ \bibinfo {author} {\bibfnamefont {F.}~\bibnamefont {Mila}},\
  }\bibfield  {title} {\enquote {\bibinfo {title} {Floating phase versus chiral
  transition in a 1d hard-boson model},}\ }\href {\doibase
  10.1103/PhysRevLett.122.017205} {\bibfield  {journal} {\bibinfo  {journal}
  {Phys. Rev. Lett.}\ }\textbf {\bibinfo {volume} {122}},\ \bibinfo {pages}
  {017205} (\bibinfo {year} {2019})}\BibitemShut {NoStop}%
\bibitem [{\citenamefont {{Turner}}\ \emph {et~al.}(2018)\citenamefont
  {{Turner}}, \citenamefont {{Michailidis}}, \citenamefont {{Abanin}},
  \citenamefont {{Serbyn}}, \citenamefont {{Papi{\'c}}},\ and\ \citenamefont
  {{}}}]{turner2018}%
  \BibitemOpen
  \bibfield  {author} {\bibinfo {author} {\bibfnamefont {C.~J.}\ \bibnamefont
  {{Turner}}}, \bibinfo {author} {\bibfnamefont {A.~A.}\ \bibnamefont
  {{Michailidis}}}, \bibinfo {author} {\bibfnamefont {D.~A.}\ \bibnamefont
  {{Abanin}}}, \bibinfo {author} {\bibfnamefont {M.}~\bibnamefont {{Serbyn}}},
  \bibinfo {author} {\bibnamefont {{Papi{\'c}}}}, \ and\ \bibinfo {author}
  {\bibfnamefont {Z.}~\bibnamefont {{}}},\ }\bibfield  {title} {\enquote
  {\bibinfo {title} {{Weak ergodicity breaking from quantum many-body
  scars}},}\ }\href {\doibase 10.1038/s41567-018-0137-5} {\bibfield  {journal}
  {\bibinfo  {journal} {Nat. Phys.}\ }\textbf {\bibinfo {volume} {14}},\
  \bibinfo {pages} {745} (\bibinfo {year} {2018})}\BibitemShut {NoStop}%
\bibitem [{\citenamefont {Ho}\ \emph {et~al.}(2019)\citenamefont {Ho},
  \citenamefont {Choi}, \citenamefont {Pichler},\ and\ \citenamefont
  {Lukin}}]{wenwei2019}%
  \BibitemOpen
  \bibfield  {author} {\bibinfo {author} {\bibfnamefont {W.~W.}\ \bibnamefont
  {Ho}}, \bibinfo {author} {\bibfnamefont {S.}~\bibnamefont {Choi}}, \bibinfo
  {author} {\bibfnamefont {H.}~\bibnamefont {Pichler}}, \ and\ \bibinfo
  {author} {\bibfnamefont {M.~D.}\ \bibnamefont {Lukin}},\ }\bibfield  {title}
  {\enquote {\bibinfo {title} {Periodic orbits, entanglement, and quantum
  many-body scars in constrained models: Matrix product state approach},}\
  }\href {\doibase 10.1103/PhysRevLett.122.040603} {\bibfield  {journal}
  {\bibinfo  {journal} {Phys. Rev. Lett.}\ }\textbf {\bibinfo {volume} {122}},\
  \bibinfo {pages} {040603} (\bibinfo {year} {2019})}\BibitemShut {NoStop}%
\bibitem [{\citenamefont {Whitsitt}\ \emph {et~al.}(2018)\citenamefont
  {Whitsitt}, \citenamefont {Samajdar},\ and\ \citenamefont
  {Sachdev}}]{Seth2018}%
  \BibitemOpen
  \bibfield  {author} {\bibinfo {author} {\bibfnamefont {S.}~\bibnamefont
  {Whitsitt}}, \bibinfo {author} {\bibfnamefont {R.}~\bibnamefont {Samajdar}},
  \ and\ \bibinfo {author} {\bibfnamefont {S.}~\bibnamefont {Sachdev}},\
  }\bibfield  {title} {\enquote {\bibinfo {title} {Quantum field theory for the
  chiral clock transition in one spatial dimension},}\ }\href {\doibase
  10.1103/PhysRevB.98.205118} {\bibfield  {journal} {\bibinfo  {journal} {Phys.
  Rev. B}\ }\textbf {\bibinfo {volume} {98}},\ \bibinfo {pages} {205118}
  (\bibinfo {year} {2018})}\BibitemShut {NoStop}%
\bibitem [{\citenamefont {Ostlund}(1981)}]{ostlund1981incommensurate}%
  \BibitemOpen
  \bibfield  {author} {\bibinfo {author} {\bibfnamefont {S.}~\bibnamefont
  {Ostlund}},\ }\bibfield  {title} {\enquote {\bibinfo {title} {Incommensurate
  and commensurate phases in asymmetric clock models},}\ }\href {\doibase
  10.1103/PhysRevB.24.398} {\bibfield  {journal} {\bibinfo  {journal} {Phys.
  Rev. B}\ }\textbf {\bibinfo {volume} {24}},\ \bibinfo {pages} {398} (\bibinfo
  {year} {1981})}\BibitemShut {NoStop}%
\bibitem [{\citenamefont {Huse}\ and\ \citenamefont
  {Fisher}(1982)}]{huse1982domain}%
  \BibitemOpen
  \bibfield  {author} {\bibinfo {author} {\bibfnamefont {D.~A.}\ \bibnamefont
  {Huse}}\ and\ \bibinfo {author} {\bibfnamefont {M.~E.}\ \bibnamefont
  {Fisher}},\ }\bibfield  {title} {\enquote {\bibinfo {title} {Domain walls and
  the melting of commensurate surface phases},}\ }\href {\doibase
  10.1103/PhysRevLett.49.793} {\bibfield  {journal} {\bibinfo  {journal} {Phys.
  Rev. Lett.}\ }\textbf {\bibinfo {volume} {49}},\ \bibinfo {pages} {793}
  (\bibinfo {year} {1982})}\BibitemShut {NoStop}%
\bibitem [{\citenamefont {Zamolodchikov}(1989)}]{ZAM1989}%
  \BibitemOpen
  \bibfield  {author} {\bibinfo {author} {\bibfnamefont {A.~B.}\ \bibnamefont
  {Zamolodchikov}},\ }\bibfield  {title} {\enquote {\bibinfo {title} {Integrals
  of motion and s-matrix of the (scaled) t = tc ising model with magnetic
  field},}\ }\href {\doibase 10.1142/S0217751X8900176X} {\bibfield  {journal}
  {\bibinfo  {journal} {Int. J. Mod. Phys. A}\ }\textbf {\bibinfo {volume}
  {04}},\ \bibinfo {pages} {4235} (\bibinfo {year} {1989})}\BibitemShut
  {NoStop}%
\bibitem [{\citenamefont {Delfino}\ and\ \citenamefont
  {Grinza}(2008)}]{delfino2008}%
  \BibitemOpen
  \bibfield  {author} {\bibinfo {author} {\bibfnamefont {G.}~\bibnamefont
  {Delfino}}\ and\ \bibinfo {author} {\bibfnamefont {P.}~\bibnamefont
  {Grinza}},\ }\bibfield  {title} {\enquote {\bibinfo {title} {Confinement in
  the q-state potts field theory},}\ }\href {\doibase
  https://doi.org/10.1016/j.nuclphysb.2007.09.003} {\bibfield  {journal}
  {\bibinfo  {journal} {Nucl. Phys. B}\ }\textbf {\bibinfo {volume} {791}},\
  \bibinfo {pages} {265} (\bibinfo {year} {2008})}\BibitemShut {NoStop}%
\bibitem [{\citenamefont {{Lepori}}\ \emph {et~al.}(2009)\citenamefont
  {{Lepori}}, \citenamefont {{T{\'o}th}},\ and\ \citenamefont
  {{Delfino}}}]{Lepori2009}%
  \BibitemOpen
  \bibfield  {author} {\bibinfo {author} {\bibfnamefont {L.}~\bibnamefont
  {{Lepori}}}, \bibinfo {author} {\bibfnamefont {G.~Z.}\ \bibnamefont
  {{T{\'o}th}}}, \ and\ \bibinfo {author} {\bibfnamefont {G.}~\bibnamefont
  {{Delfino}}},\ }\bibfield  {title} {\enquote {\bibinfo {title} {{The particle
  spectrum of the three-state Potts field theory: a numerical study}},}\ }\href
  {\doibase 10.1088/1742-5468/2009/11/P11007} {\bibfield  {journal} {\bibinfo
  {journal} {J. Stat. Mech.: Theory Exp}\ }\textbf {\bibinfo {volume} {2009}},\
  \bibinfo {pages} {11007} (\bibinfo {year} {2009})}\BibitemShut {NoStop}%
\bibitem [{\citenamefont {{Rutkevich}}(2015)}]{Rutkevich2015}%
  \BibitemOpen
  \bibfield  {author} {\bibinfo {author} {\bibfnamefont {S.~B.}\ \bibnamefont
  {{Rutkevich}}},\ }\bibfield  {title} {\enquote {\bibinfo {title} {{Baryon
  masses in the three-state Potts field theory in a weak magnetic field}},}\
  }\href {\doibase 10.1088/1742-5468/2015/01/P01010} {\bibfield  {journal}
  {\bibinfo  {journal} {J. Stat. Mech.: Theory Exp}\ }\textbf {\bibinfo
  {volume} {2015}},\ \bibinfo {eid} {01010} (\bibinfo {year}
  {2015})}\BibitemShut {NoStop}%
\bibitem [{\citenamefont {{Lencs{\'e}s}}\ and\ \citenamefont
  {{Tak{\'a}cs}}(2015)}]{lencses2015}%
  \BibitemOpen
  \bibfield  {author} {\bibinfo {author} {\bibfnamefont {M.}~\bibnamefont
  {{Lencs{\'e}s}}}\ and\ \bibinfo {author} {\bibfnamefont {G.}~\bibnamefont
  {{Tak{\'a}cs}}},\ }\bibfield  {title} {\enquote {\bibinfo {title}
  {{Confinement in the q-state Potts model: an RG-TCSA study}},}\ }\href
  {\doibase 10.1007/JHEP09(2015)146} {\bibfield  {journal} {\bibinfo  {journal}
  {J. High Energy Phys}\ }\textbf {\bibinfo {volume} {2015}},\ \bibinfo {eid}
  {146} (\bibinfo {year} {2015})}\BibitemShut {NoStop}%
\bibitem [{\citenamefont {Omran}\ \emph {et~al.}(2019)\citenamefont {Omran},
  \citenamefont {Levine}, \citenamefont {Keesling}, \citenamefont {Semeghini},
  \citenamefont {Wang}, \citenamefont {Ebadi}, \citenamefont {Bernien},
  \citenamefont {Zibrov}, \citenamefont {Pichler}, \citenamefont {Choi},
  \citenamefont {Cui}, \citenamefont {Rossignolo}, \citenamefont {Rembold},
  \citenamefont {Montangero}, \citenamefont {Calarco}, \citenamefont {Endres},
  \citenamefont {Greiner}, \citenamefont {Vuleti{\'c}},\ and\ \citenamefont
  {Lukin}}]{Omran570}%
  \BibitemOpen
  \bibfield  {author} {\bibinfo {author} {\bibfnamefont {A.}~\bibnamefont
  {Omran}}, \bibinfo {author} {\bibfnamefont {H.}~\bibnamefont {Levine}},
  \bibinfo {author} {\bibfnamefont {A.}~\bibnamefont {Keesling}}, \bibinfo
  {author} {\bibfnamefont {G.}~\bibnamefont {Semeghini}}, \bibinfo {author}
  {\bibfnamefont {T.~T.}\ \bibnamefont {Wang}}, \bibinfo {author}
  {\bibfnamefont {S.}~\bibnamefont {Ebadi}}, \bibinfo {author} {\bibfnamefont
  {H.}~\bibnamefont {Bernien}}, \bibinfo {author} {\bibfnamefont {A.~S.}\
  \bibnamefont {Zibrov}}, \bibinfo {author} {\bibfnamefont {H.}~\bibnamefont
  {Pichler}}, \bibinfo {author} {\bibfnamefont {S.}~\bibnamefont {Choi}},
  \bibinfo {author} {\bibfnamefont {J.}~\bibnamefont {Cui}}, \bibinfo {author}
  {\bibfnamefont {M.}~\bibnamefont {Rossignolo}}, \bibinfo {author}
  {\bibfnamefont {P.}~\bibnamefont {Rembold}}, \bibinfo {author} {\bibfnamefont
  {S.}~\bibnamefont {Montangero}}, \bibinfo {author} {\bibfnamefont
  {T.}~\bibnamefont {Calarco}}, \bibinfo {author} {\bibfnamefont
  {M.}~\bibnamefont {Endres}}, \bibinfo {author} {\bibfnamefont
  {M.}~\bibnamefont {Greiner}}, \bibinfo {author} {\bibfnamefont
  {V.}~\bibnamefont {Vuleti{\'c}}}, \ and\ \bibinfo {author} {\bibfnamefont
  {M.~D.}\ \bibnamefont {Lukin}},\ }\bibfield  {title} {\enquote {\bibinfo
  {title} {Generation and manipulation of schr{\"o}dinger cat states in rydberg
  atom arrays},}\ }\href {\doibase 10.1126/science.aax9743} {\bibfield
  {journal} {\bibinfo  {journal} {Science}\ }\textbf {\bibinfo {volume}
  {365}},\ \bibinfo {pages} {570} (\bibinfo {year} {2019})}\BibitemShut
  {NoStop}%
\bibitem [{\citenamefont {Elitzur}\ \emph {et~al.}(1979)\citenamefont
  {Elitzur}, \citenamefont {Pearson},\ and\ \citenamefont
  {Shigemitsu}}]{elitzur1979}%
  \BibitemOpen
  \bibfield  {author} {\bibinfo {author} {\bibfnamefont {S.}~\bibnamefont
  {Elitzur}}, \bibinfo {author} {\bibfnamefont {R.~B.}\ \bibnamefont
  {Pearson}}, \ and\ \bibinfo {author} {\bibfnamefont {J.}~\bibnamefont
  {Shigemitsu}},\ }\bibfield  {title} {\enquote {\bibinfo {title} {Phase
  structure of discrete abelian spin and gauge systems},}\ }\href {\doibase
  10.1103/PhysRevD.19.3698} {\bibfield  {journal} {\bibinfo  {journal} {Phys.
  Rev. D}\ }\textbf {\bibinfo {volume} {19}},\ \bibinfo {pages} {3698}
  (\bibinfo {year} {1979})}\BibitemShut {NoStop}%
\bibitem [{\citenamefont {Einhorn}\ \emph {et~al.}(1980)\citenamefont
  {Einhorn}, \citenamefont {Savit},\ and\ \citenamefont
  {Rabinovici}}]{einhorn1980}%
  \BibitemOpen
  \bibfield  {author} {\bibinfo {author} {\bibfnamefont {M.~B.}\ \bibnamefont
  {Einhorn}}, \bibinfo {author} {\bibfnamefont {R.}~\bibnamefont {Savit}}, \
  and\ \bibinfo {author} {\bibfnamefont {E.}~\bibnamefont {Rabinovici}},\
  }\bibfield  {title} {\enquote {\bibinfo {title} {A physical picture for the
  phase transitions in zn symmetric models},}\ }\href {\doibase
  https://doi.org/10.1016/0550-3213(80)90473-3} {\bibfield  {journal} {\bibinfo
   {journal} {Nucl. Phys. B}\ }\textbf {\bibinfo {volume} {170}},\ \bibinfo
  {pages} {16} (\bibinfo {year} {1980})}\BibitemShut {NoStop}%
\bibitem [{\citenamefont {Coleman}\ \emph {et~al.}(1975)\citenamefont
  {Coleman}, \citenamefont {Jackiw},\ and\ \citenamefont
  {Susskind}}]{coleman1975}%
  \BibitemOpen
  \bibfield  {author} {\bibinfo {author} {\bibfnamefont {S.}~\bibnamefont
  {Coleman}}, \bibinfo {author} {\bibfnamefont {R.}~\bibnamefont {Jackiw}}, \
  and\ \bibinfo {author} {\bibfnamefont {L.}~\bibnamefont {Susskind}},\
  }\bibfield  {title} {\enquote {\bibinfo {title} {Charge shielding and quark
  confinement in the massive schwinger model},}\ }\href {\doibase
  https://doi.org/10.1016/0003-4916(75)90212-2} {\bibfield  {journal} {\bibinfo
   {journal} {Ann. Phys.}\ }\textbf {\bibinfo {volume} {93}},\ \bibinfo {pages}
  {267} (\bibinfo {year} {1975})}\BibitemShut {NoStop}%
\bibitem [{\citenamefont {Coleman}(1976)}]{coleman1976}%
  \BibitemOpen
  \bibfield  {author} {\bibinfo {author} {\bibfnamefont {S.}~\bibnamefont
  {Coleman}},\ }\bibfield  {title} {\enquote {\bibinfo {title} {More about the
  massive schwinger model},}\ }\href {\doibase
  https://doi.org/10.1016/0003-4916(76)90280-3} {\bibfield  {journal} {\bibinfo
   {journal} {Ann. Phys.}\ }\textbf {\bibinfo {volume} {101}},\ \bibinfo
  {pages} {239} (\bibinfo {year} {1976})}\BibitemShut {NoStop}%
\bibitem [{\citenamefont {Delfino}\ and\ \citenamefont
  {Mussardo}(1998)}]{multifreqSG}%
  \BibitemOpen
  \bibfield  {author} {\bibinfo {author} {\bibfnamefont {G.}~\bibnamefont
  {Delfino}}\ and\ \bibinfo {author} {\bibfnamefont {G.}~\bibnamefont
  {Mussardo}},\ }\bibfield  {title} {\enquote {\bibinfo {title} {Non-integrable
  aspects of the multi-frequency sine-gordon model},}\ }\href {\doibase
  https://doi.org/10.1016/S0550-3213(98)00063-7} {\bibfield  {journal}
  {\bibinfo  {journal} {Nucl. Phys. B}\ }\textbf {\bibinfo {volume} {516}},\
  \bibinfo {pages} {675} (\bibinfo {year} {1998})}\BibitemShut {NoStop}%
\bibitem [{\citenamefont {Shankar}\ and\ \citenamefont
  {Murthy}(2005)}]{Shankar05}%
  \BibitemOpen
  \bibfield  {author} {\bibinfo {author} {\bibfnamefont {R.}~\bibnamefont
  {Shankar}}\ and\ \bibinfo {author} {\bibfnamefont {G.}~\bibnamefont
  {Murthy}},\ }\bibfield  {title} {\enquote {\bibinfo {title} {Deconfinement in
  $d=1$: Asymptotic and half-asymptotic particles},}\ }\href {\doibase
  10.1103/PhysRevB.72.224414} {\bibfield  {journal} {\bibinfo  {journal} {Phys.
  Rev. B}\ }\textbf {\bibinfo {volume} {72}},\ \bibinfo {pages} {224414}
  (\bibinfo {year} {2005})}\BibitemShut {NoStop}%
\bibitem [{Note2()}]{Note2}%
  \BibitemOpen
  \bibinfo {note} {The transition from the vacuum state to the mesonic
  (baryonic) excitation involves a second (third) order process.}\BibitemShut
  {Stop}%
\bibitem [{Note3()}]{Note3}%
  \BibitemOpen
  \bibinfo {note} {The `...' denotes repeated structure of the ordered phase
  with rgg}\BibitemShut {NoStop}%
\bibitem [{\citenamefont {Yang}\ \emph {et~al.}(2020)\citenamefont {Yang},
  \citenamefont {Liu}, \citenamefont {Gorshkov},\ and\ \citenamefont
  {Iadecola}}]{zhicheng2019}%
  \BibitemOpen
  \bibfield  {author} {\bibinfo {author} {\bibfnamefont {Z.-C.}\ \bibnamefont
  {Yang}}, \bibinfo {author} {\bibfnamefont {F.}~\bibnamefont {Liu}}, \bibinfo
  {author} {\bibfnamefont {A.~V.}\ \bibnamefont {Gorshkov}}, \ and\ \bibinfo
  {author} {\bibfnamefont {T.}~\bibnamefont {Iadecola}},\ }\bibfield  {title}
  {\enquote {\bibinfo {title} {Hilbert-space fragmentation from strict
  confinement},}\ }\href {\doibase 10.1103/PhysRevLett.124.207602} {\bibfield
  {journal} {\bibinfo  {journal} {Phys. Rev. Lett.}\ }\textbf {\bibinfo
  {volume} {124}},\ \bibinfo {pages} {207602} (\bibinfo {year}
  {2020})}\BibitemShut {NoStop}%
\bibitem [{Note4()}]{Note4}%
  \BibitemOpen
  \bibinfo {note} {The light-cone shows a $\protect \mathbb {Z}_3$-periodic
  sub-structure due to the blockade physics.}\BibitemShut {Stop}%
\bibitem [{Note5()}]{Note5}%
  \BibitemOpen
  \bibinfo {note} {The local rotation fields can be also engineered by applying
  a strong light shift to the selected target atoms, and shining an additional
  resonant Rydberg laser beam across the whole chain.}\BibitemShut {Stop}%
\bibitem [{\citenamefont {Samajdar}\ \emph {et~al.}(2020)\citenamefont
  {Samajdar}, \citenamefont {Ho}, \citenamefont {Pichler}, \citenamefont
  {Lukin},\ and\ \citenamefont {Sachdev}}]{Samajdar2020}%
  \BibitemOpen
  \bibfield  {author} {\bibinfo {author} {\bibfnamefont {R.}~\bibnamefont
  {Samajdar}}, \bibinfo {author} {\bibfnamefont {W.~W.}\ \bibnamefont {Ho}},
  \bibinfo {author} {\bibfnamefont {H.}~\bibnamefont {Pichler}}, \bibinfo
  {author} {\bibfnamefont {M.~D.}\ \bibnamefont {Lukin}}, \ and\ \bibinfo
  {author} {\bibfnamefont {S.}~\bibnamefont {Sachdev}},\ }\bibfield  {title}
  {\enquote {\bibinfo {title} {Complex density wave orders and quantum phase
  transitions in a model of square-lattice rydberg atom arrays},}\ }\href
  {\doibase 10.1103/PhysRevLett.124.103601} {\bibfield  {journal} {\bibinfo
  {journal} {Phys. Rev. Lett.}\ }\textbf {\bibinfo {volume} {124}},\ \bibinfo
  {pages} {103601} (\bibinfo {year} {2020})}\BibitemShut {NoStop}%
\bibitem [{\citenamefont {Rapp}\ \emph {et~al.}(2007)\citenamefont {Rapp},
  \citenamefont {Zar{\'{a}}nd}, \citenamefont {Honerkamp},\ and\ \citenamefont
  {Hofstetter}}]{Rapp2007}%
  \BibitemOpen
  \bibfield  {author} {\bibinfo {author} {\bibfnamefont {{\'{A}}.}~\bibnamefont
  {Rapp}}, \bibinfo {author} {\bibfnamefont {G.}~\bibnamefont {Zar{\'{a}}nd}},
  \bibinfo {author} {\bibfnamefont {C.}~\bibnamefont {Honerkamp}}, \ and\
  \bibinfo {author} {\bibfnamefont {W.}~\bibnamefont {Hofstetter}},\ }\bibfield
   {title} {\enquote {\bibinfo {title} {{Color superfluidity and "baryon"
  formation in ultracold fermions}},}\ }\href {\doibase
  10.1103/PhysRevLett.98.160405} {\bibfield  {journal} {\bibinfo  {journal}
  {Phys. Rev. Lett.}\ }\textbf {\bibinfo {volume} {98}},\ \bibinfo {pages} {1}
  (\bibinfo {year} {2007})}\BibitemShut {NoStop}%
\end{thebibliography}%

\clearpage
\newpage 
\onecolumngrid
\setcounter{figure}{0}
\makeatletter
\renewcommand{\thefigure}{S\@arabic\c@figure}
\setcounter{equation}{0} \makeatletter
\renewcommand \theequation{S\@arabic\c@equation}
\renewcommand \thetable{S\@arabic\c@table}

\end{document}